# Theory of high-energy messengers

**Charles D. Dermer**

Code 7653, US Naval Research Laboratory, 4555 Overlook Ave. SW, Washington, DC 20375 USA[1]

Charles.Dermer@outlook.com, @chuckdermer

**Abstract**. Knowledge of the distant high-energy universe comes from photons, ultra-high energy cosmic rays (UHECRs), high-energy neutrinos, and gravitational waves. The theory of high-energy messengers reviewed here focuses on the extragalactic background light at all wavelengths, cosmic rays and magnetic fields in intergalactic space, and neutrinos of extragalactic origin. Comparisons are drawn between the intensities of photons and UHECRs in intergalactic space, and the high-energy neutrinos recently detected with IceCube at about the Waxman-Bahcall flux. Source candidates for UHECRs and high-energy neutrinos are reviewed, focusing on star-forming and radio-loud active galaxies. HAWC and Advanced LIGO are just underway, with much anticipation.

## 1. Introduction

One of the extraordinary scientific achievements of the past 10 decade is IceCube's opening of the high-energy, $> \sim$ 30 TeV – PeV neutrino sky [1,2,3]. With increased exposure, improved analyses, and larger and better neutrino detectors, we may ultimately trace these neutrinos to their sources. The still relatively small total number , ~50, of events above the cosmic-ray induced neutrino flux, particle physics uncertainties, and the small fraction, ~25%, of events that are the better-resolved tracks means that association with members of any given source class is yet statistically very weak. Hoped-for temporal coincidences of a high-energy neutrino with a flaring source could however easily flip the odds in favor of a case for association, and is actively sought.

Another extraordinary scientific development during the past 10 years, though in the opposite sense, is the disappearance, or at least dramatic fading of the anisotropy in the arrival direction distribution of UHECRs observed with the Pierre Auger Observatory (PAO). What was first announced in November 2007 as a 99% confidence anisotropy for UHECRs with energy $E > 6 \times 10^{19}$ eV [4] correlated with active galactic nuclei (AGN) in the Véron-Cetty & Véron catalog [5] that trace the supergalactic plane within 75 Mpc has, like the Cheshire Cat, vanished except for a suggestive grin in the direction of Cen A [6]. The birth of cosmic-ray astronomy, where particles can be traced to their sources, remains as elusive as ever, even taking into account the Telescope Array (TA) hotspot of arriving UHECRs with E > 57 EeV [7], notably in a different direction (RA = 146.7°, decl. = 43.2°) than an earlier hotspot indicated by Haverah Park, AGASA, Volcano Ranch and Yakutsk data towards the supergalactic plane around M87 (RA = 187.7°, decl. = 12.4°) [8]. Fortunately, conclusions from PAO and HiRes regarding UHECR composition at the highest energies seem to be converging on a mixed composition above $\approx (0.3 - 1) \times 10^{19}$ eV.

Other big news in multi-messenger astronomy is Advanced LIGO, now switched on and operating since 18 September 2015 at design sensitivity. The groundbreaking a-LIGO announcement of gravitational waves from colliding black holes has thrilled the astronomical community by opening the gravitational wave window, and more discoveries are anticipated as a-VIRGO joins the search within the year. Other news in high-energy multi-messenger astronomy is the July 2015 inauguration of the large field-of-view TeV γ-ray telescope HAWC with its 300 tanks high on the slopes of Sierra Negra near Puebla, Mexico. HAWC promises a unique view of the TeV sky to complement pointing, ground-based VHE (>~ 50 – 100 GeV) γ-ray telescopes. The VERITAS 4x12 m telescope array, the Magic 2x17 m array, and the HESS-II 4x12 m + 1x27 m array keep producing extraordinary results, for example, rapid, few minute variability of BL Lac objects, 10 min variability of flat (radio) spectrum radio quasars (FSRQs), and now VHE γ-ray detections of 5 distant FSRQ blazars, most recently, PKS 1441+25 at redshift z = 0.94 [9]. The siting decision of the Cherenkov Telescope Array (CTA), with the major Southern Hemisphere array in Chile, and the smaller Northern Hemisphere array in the Canary Islands, was announced July 2015, and construction and early science operations could be expected as early as 2018, with completion by 2020 – 2021.

The backbone of high-energy astronomy remains the Fermi Gamma ray Space Telescope, tirelessly accumulating data on the whole sky since its launch in 2008, and prepared to be a trigger for high-energy neutrino, gravitational wave, and even UHECR events. It and the Swift Gamma Ray Burst (GRB) Mission launched in 2004 have been vastly successful in advancing GRB and AGN/blazar science.

Some features of the high-energy sky at the birth of the multi-messenger era are summarized in this TAUP-Torino 2015 paper, including one of the great problems of high-energy astronomy: the origin of the UHECRs. This is a problem that has been with us since Bruno Rossi wrote in 1934 that "…occasionally very extensive groups of particles arrive upon the experiment," and Pierre Auger and collaborators undertook a dedicated study of air-shower coincidences in 1938 [10]. Special attention is given to blazars—active galaxies with jets pointed in our direction—because they are fantastical objects that are candidate UHECR sources and furthermore make up a large fraction of the extragalactic γ-ray background (EGB) light. It—the EGB—is part of the extragalactic background light (EBL), a subject which commences this paper.

## 2. The extragalactic night sky

Most of us are aware that Olbers' paradox is resolved by the fact of our finite, evolving universe. A universe with a finite lifetime, or one that expands and cools forever, dethrones any hope for an infinite non-expanding steady-state universe. Any existential anxiety produced by this fact may or may not be relieved by knowing that a universe with our specific ΛCDM cosmology allows life forms of our Darwinian kind to exist.

Given that we are here, and that the night sky is dark, we can ask what information is encoded in the intensity spectrum of the EBL at some typical randomly chosen spot in the local (redshift $z \ll 1$) universe, far from the Sun and the zodiacal light that contaminates the night sky flux as a result of Sunlight scattered by Solar System dust, and far from the Milky Way which adds additional uncertainty from, primarily, the model-dependent intensity of the isotropic part of the cosmic-ray electron synchrotron flux. These fields complicate the actual determination of the EBL, and γ-ray astronomy provides ways to constrain the EBL intensity, e.g., by observing γ-ray opacity cutoffs in the spectral energy distributions (SEDs) of high-energy γ-ray sources at various cosmological distances [11]. This effect, due to the $\gamma + \gamma \rightarrow e^+ + e^-$ process, is a consequence of γ rays interacting with target IR/optical EBL photons during their transit through the universe from source to observer.

*2.1. The EBL*

Fig. 1 is a plot of the intensity of the EBL (noting that this acronym is sometimes used specifically for the IR/optical/UV portion of the EBL) and the isotropic cosmic-ray and UHECR intensity at the current epoch ($z \rightarrow 0$), written in the form of an energy density, from my book [12] with Govind

Menon published in 2009 but based on pre-Fermi data. Because the lower-energy cosmic rays are Galactic and so would not be measured in *inter*galactic space, they are formally out of place here (though they do illustrate the coincidence between the values of the Galactic cosmic-ray and CMBR energy densities). The actual transition energy from Galactic Cosmic Rays to Extragalactic UHECRs is however unknown, ranging from at least as low as the second knee at $\sim 2\times10^{17}$ eV in the cosmic ray spectrum, to as high as the ankle at $\sim 3\times10^{18}$ eV (assuming that the highest-energy cosmic rays are indeed extragalactic).

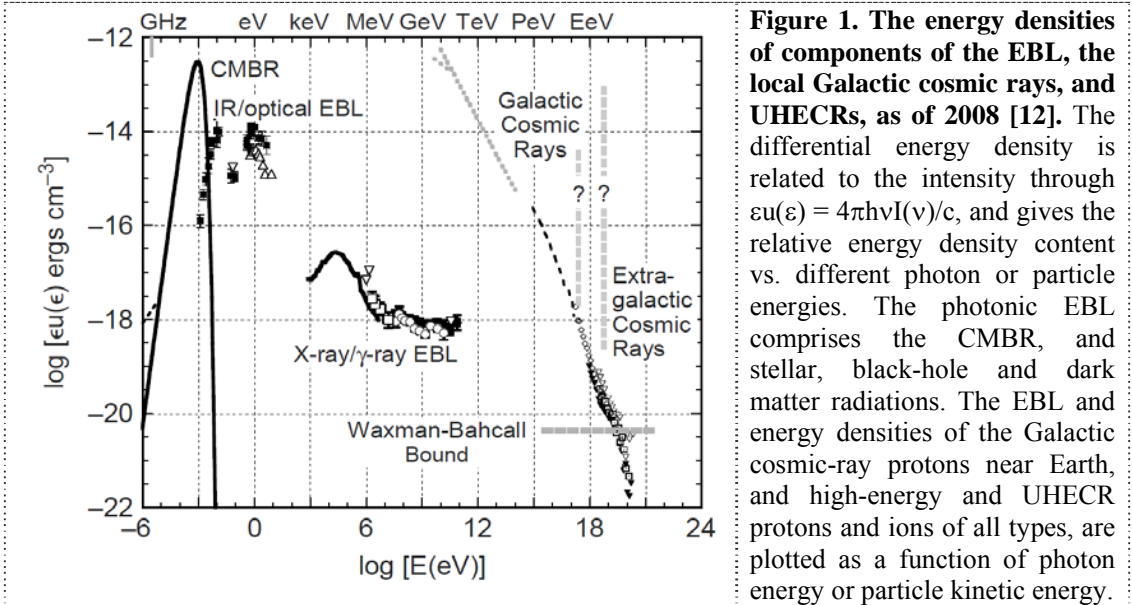

**Figure 1. The energy densities of components of the EBL, the local Galactic cosmic rays, and UHECRs, as of 2008 [12].** The differential energy density is related to the intensity through $\varepsilon u(\varepsilon) = 4\pi h\nu I(\nu)/c$, and gives the relative energy density content vs. different photon or particle energies. The photonic EBL comprises the CMBR, and stellar, black-hole and dark matter radiations. The EBL and energy densities of the Galactic cosmic-ray protons near Earth, and high-energy and UHECR protons and ions of all types, are plotted as a function of photon energy or particle kinetic energy.

The dominant component of the EBL is the cosmic microwave background radiation (CMBR). The amazing precision cosmology obtained most recently through analyses of Planck angular power spectra data, as reviewed at this conference by Dr. A. Challinor, must be acknowledged and applauded. Here, though, we consider the CMBR as (indeed) it is, the cooling remnant of the big bang, an important background radiation field to be sure, but a field distinct from those fields related to structure formation processes after decoupling that led to stars and galaxies and creatures like the attendees of TAUP 2015.

Stellar processes from all classes of galaxies, but mainly from bluish star-forming galaxies, make the optical peak at $\sim 1\mu$, and stellar emissions reprocessed from gaseous and dusty regions mainly near hot young stars are primarily responsible for the IR peak at $\sim 100 - 200\mu$. Radio-quiet AGNs, including Type 1 Seyfert galaxies with small column densities of directly obscuring gas, though with broad line region (BLR) clouds and a dusty tours, and gas-enshrouded AGN beasts surrounded by Thomson-thick columns of gas or BLR clouds with large covering fractions, make almost all of the X-ray EBL. The origin of the $>\sim 100$ MeV $\gamma$-ray EBL is surely dominated by blazars, though it has required Fermi to quantify this fraction, as the Compton Gamma Ray Observatory EGRET data shown in Fig. 1 suffered from a calibration underestimation of EGRET's effective area above $\approx 1$ GeV.

The EBL intensity, like the UHECR intensity, though with importantly different energy-loss and production channels, gives an integrated, bolometric description of all the normal-matter and dark-matter activity throughout the history of the universe, and so constrains the stellar and black-hole activity of any model of the universe. The ratios of energy densities in the IR/optical, X-ray and >> MeV $\gamma$ ray EBLs to the CMBR energy density are $\approx 0.05$, $5\times 10^{-4}$, and $10^{-5}$, respectively, reflecting the aggregate energy generation by stars, radio-quiet AGNs and radio galaxies and blazars, respectively, since the age of the first stars.

Note how structured the photonic EBL intensity is by comparison with the comparatively smooth >> 1 GeV/nucleon cosmic-ray flux, which itself must be composed of at least 2 and probably 3 components. This curiosity, if it is indeed so, lacks a compelling explanation known to the author.

*2.2. The transparent universe and the Waxman-Bahcall flux*

Major upgrades have been made to our knowledge of radiation fields in extragalactic space in the last 10 years, the most important being the Auger [13] and High Resolution Fly's Eye (HiRes) [14] detection of the Greisen-Zatsepin-Kuzmin (GZK) cutoff in the UHECR spectrum at ≈ 6×10$^{19}$ eV (or at least a cutoff coincident with this energy), the IceCube detection of extragalactic neutrinos [1]-[3], and the unveiling of the Fermi-Large Area Telescope (LAT) GeV sky [15]. See Fig. 2.

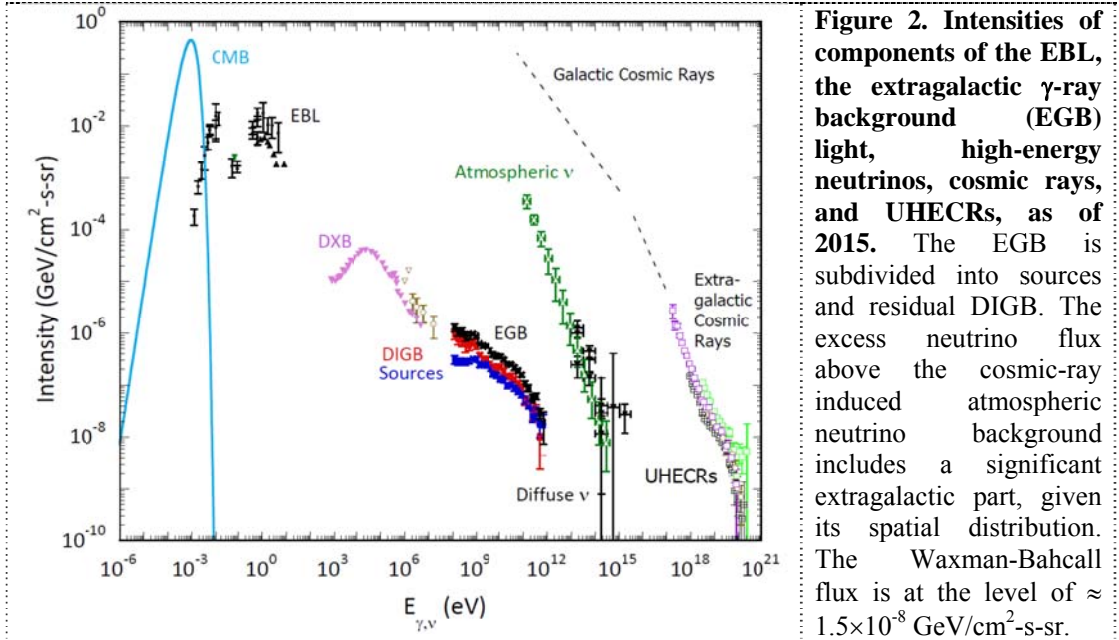

**Figure 2. Intensities of components of the EBL, the extragalactic γ-ray background (EGB) light, high-energy neutrinos, cosmic rays, and UHECRs, as of 2015.** The EGB is subdivided into sources and residual DIGB. The excess neutrino flux above the cosmic-ray induced atmospheric neutrino background includes a significant extragalactic part, given its spatial distribution. The Waxman-Bahcall flux is at the level of ≈ 1.5×10$^{-8}$ GeV/cm$^2$-s-sr.

High-energy neutrinos are notoriously immune to interactions: it takes the grammage of a chord of material through the Earth to stop them at >> PeV energies. By comparison, γ rays are not so elusive, but those between ~1 MeV and 20 GeV pass through the z <~ 10 universe generally unscathed except for the redshift effect. UHECR protons, on the other hand, suffer severe energy losses above the GZK energy, due (by definition) to photopion production by UHECR protons and photodisintegration of UHECR ions with CMB photons. The low-z CMBR photon number density $n_{CMBR}$ ≈ 400 cm$^{-3}$, implying an asymptotic energy-loss length of $\lambda_{GZK}$ = (70 μb × $n_{CMBR}$)$^{-1}$ ≈ 12 Mpc energy for UHECR protons with energy $E_p$ ≈ $m_p$(2 $m_\pi/m_e$) / ⟨$\varepsilon_{CMBR}$⟩ ~ 4×10$^{20}$ eV making photopions by interacting with CMB photons, noting that the mean photon energy of the z << 1 CMBR is ⟨$\varepsilon_{CMBR}$⟩ ≈ 10$^{-9}$ in $m_e$ units. The photopion energy-loss length of 10$^{20}$ eV UHECR protons is ≈ 140 Mpc [16].

For all intents and purposes, neutrinos (ν) and GeV γ rays travel through intergalactic space unhindered, so the equation of radiative transfer, $dI_\varepsilon/ds$ = -$\kappa_\varepsilon I_\varepsilon$ + j(ε,Ω) with spectral absorption coefficient $\kappa_\varepsilon$ → 0 implies that the integral γ or ν intensity I = $\zeta R_H$ j/4π, where j = $n_s$ ⟨ $L_s$ ⟩ (units of energy density over time) is the γ or ν emissivity written as the product of the source density $n_s$ and average source luminosity ⟨ $L_s$ ⟩. The Hubble Radius $R_H$ = c/$H_0$ ≅ 4300 Mpc, and ζ ~ 1 is a correction factor for cosmological effects and increased activity at earlier epochs z ~ 1 – 3. Thus, from the > 10$^{20}$ eV UHECR energy density in Fig. 1,

$$I = \varsigma \frac{R_H}{4\pi} j \approx 0.25 \varsigma \frac{R_H}{4\pi} \frac{10^{-21} erg/cm^3}{140 Mpc/c} \approx \varsigma \frac{R_H}{4\pi} \frac{10^{44} erg}{Mpc^3 - yr} \approx 1.5 \times 10^{-8} \varsigma \frac{GeV}{cm^2 - s - sr}, \quad (1)$$

using ¼ for the fraction of proton energy that goes into ν. An intensity of a few $\times 10^{-8}$ GeV/(cm$^2$ s sr) is referred to as the "Waxman-Bahcall flux" [17]. The appearance of a high-energy ν flux up to PeV energies at the Waxman-Bahcall level is predictable if some universal mechanism injects a −2 spectrum of protons and ions to the highest energies, and an efficient ν production process, like secondary nuclear production, operates calorimetrically at the lower 10 – 100 PeV energies [18].

*2.3. The EGB and DIGB*

The EGB is the extragalactic γ-ray background radiation in our local low-redshift neighborhood far from the γ-ray light of the Milky Way and its satellites such as the Small and Large Magellanic Clouds. Three sets of Fermi-LAT data for the γ-ray EBL are shown in Fig. 2 [15]. The EGB includes the unresolved diffuse intergalactic γ-ray background (DIGB) as well as emission associated with all identified source such as the bright blazars 3C 279 and 3C 454.3 and with all strongly associated extragalactic γ-ray sources. The mean intensity of all significantly associated Fermi-LAT γ-ray sources are shown in Fig. 2 by the blue data points, and the difference between the average source intensity and EBL intensity is the DIGB, shown by the red points. The DIGB could be due to a superposition of discrete sources below the Fermi sensitivity, or to a truly diffuse component, e.g., of cosmogenic origin due to cascades induced when UHECRs make photopions or photopairs.

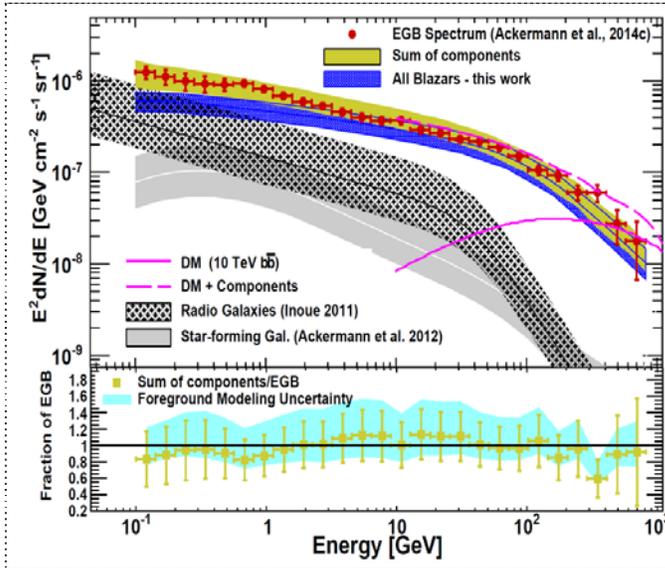

**Figure 3. Decomposition [18] of the Fermi-LAT all-sky averaged EGB intensity into separate contributions of blazars, radio galaxies [20], and star-forming galaxies [21].** The deficit at low energies from blazars can be entirely accounted for by radio and star-forming galaxy emissions. At high energies, hard γ-ray spectrum BL Lac blazars make most of the EGB, leaving only a small residual that could be made by dark matter. (Ackermann et al., 2014c = [15].) The dark-matter annihilation signal from 10 TeV $b\bar{b}$ dark matter is ruled out at the level shown.

The 100 MeV – 820 GeV EGB intensity of the EGB, separated into contributions from blazars, radio galaxies [20], and star-forming galaxies [21], is shown in Fig. 3 [19]. Blazars represent $50^{+12}_{-11}$ % of the total EGB, and Fermi-LAT has resolved 70% of the blazar contribution, implying that blazars make up < 20% of the DIGB flux. Star-forming galaxies and radio galaxies can account for most of what Fermi measures, leaving only a small residual from dark-matter annihilation or decay.

Referring back to Fig. 2, note that the EGB, source, and DIGB intensities exhibit a turndown at $E_\gamma \approx 300$ GeV (the power-law exponential cutoff solution for the EGB is best fit with a cutoff at 279 ± 52 GeV and a photon index of 2.32 ± 0.02 [15]). The low-redshift γγ pair production threshold $\varepsilon\varepsilon_1 >\sim 2$ implies that photons with energies $E_\gamma = m_e \varepsilon <\sim 2m_e \times 0.5\times 10^{-3}$ /300 ≈ 2 eV will be above the pair-production threshold for γ rays with $E_\gamma >> 300$ GeV. An energy of a few eV is where the numbers of

EBL photons from stellar processes rapidly rise, potentially explaining the cutoff in the EGB as a result of γγ absorption of (mostly) blazar γ rays by EBL photons.

If cosmogenic γ rays from reprocessed GZK (p + γ → N + π) pion-decay secondaries and cascading Bethe-Heitler (p + γ → p + e$^+$ + e$^-$) pairs from UHECRs interacting with photons of the CMBR and EBL are to make a significant fraction of the EGB, then the universal UHECR proton flux must extend well below the ankle energy [22]. UHECRs in intergalactic space, even if entirely protons, make a negligible cosmogenic contribution at the PeV energies where IceCube has reported excess neutrinos. Detection of cosmogenic neutrinos with energies >$10^{17}$ eV is more likely with the upcoming Askaryan Radio Array (ARA) [23]. The cosmogenic fluxes of γ rays and neutrinos vary considerably depending on whether the UHECRs are protons or Fe nuclei [24].

*2.4. Models for the IR/optical EBL*

Three distinct approaches have been made when modeling the IR/optical EBL. *Empirical* models directly add fluxes implied by IR/optical galaxy data at different redshifts, extrapolating to weak fluxes and high-z where the data is missing, as in the work by Stecker [25], Franceschini et al., and Dominguez et al. *A semi-analytic merger-tree model* of galaxy formation, following the coalescence of cold-dark matter halos, the onset of structure formation, and the formation of stars and galaxies, provides a second approach, found in the work by Primack, Gilmore and colleagues [26]. A model of *star formation and dust re-radiation* that directly sums the stellar and reprocessed stellar radiation is developed in the work by Kneiske et al. and Finke et al. [27].

In the latter approach, an equation for the differential stellar emissivity (in units, e.g., of erg/Mpc$^3$-yr; cf. Eq.1) of the form

$$\varepsilon j^{stars}(\varepsilon;z) = \varepsilon^2 f_{esc}(\varepsilon) \int dm\, \xi(m) \int_z^{z_{max}} dz_1 \left|\frac{dt_*}{dz_1}\right| \psi(z_1) \dot{N}_*(\varepsilon;m,t_*(z,z_1)), \quad (2)$$

is solved. The $\dot{N}_*$-term is the stellar photon production rate using stellar radii and luminosity for stars on the Hertzsprung-Russell diagram with ages when born at $z_1$ and living to z, weighted by the star-formation rate $\psi(z_1)$ (e.g., Hopkins and Beacom 2006) at birth ($z_1$). The total stellar production rate is obtained by integrating over stars of different mass m (in Solar masses), weighted by an initial mass function, e.g., of the form $\xi(m) \propto m^{-1.5}$ at 0.1 < m < 0.5, and a Salpeter IMF $\xi(m) \propto m^{-2.2}$ at m > 0.5. The fraction $f_{esc}(\varepsilon)$ of photons that escape from the Galaxy (and therefore the fraction 1− $f_{esc}(\varepsilon)$ reprocessed into the infrared given assumed dust properties) is obtained from observations of ~$10^5$ galaxies in the Millennium Galaxy survey (see [27] for references). The model assumes that $f_{esc}(\varepsilon)$ and $\xi(m)$ are z-independent, an assumption that can be relaxed in more detailed treatments.

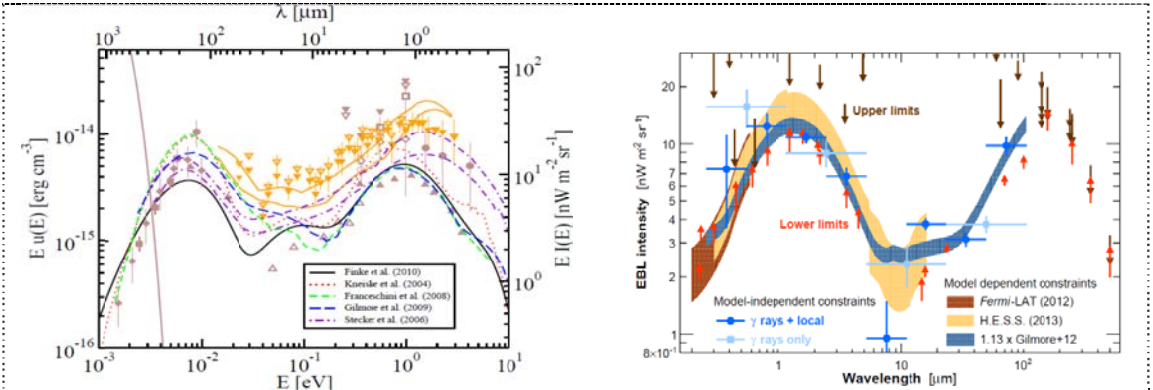

**Figure 4.** Model calculations of and model-dependent γ-ray constraints on the EBL energy density/intensity at the present epoch. *Left*: comparison of EBL energy densities from models and γ-ray constraints vs. photon energy [27]. *Right:* γ-ray constraints on the EBL, lower limits on the EBL intensity, and EBL model [26] results vs. wavelength [28].

Fig. 4, left, shows model calculations [27] and γ-ray upper-limit and observational lower-limit constraints. These constraints are updated in a recent study [28] and compared with model [26]. The lower limits are purely empirical, and involve summing all galaxies at all redshifts with low-luminosity and high-z corrections. The simplest approach to obtain joint Fermi/VHE γ-ray constraints is to extrapolate the Fermi ~1 – 10 GeV data from a TeV blazar to VHE and TeV energies, and require that the deabsorbed flux (i.e., the source flux after adding losses from γ-ray pair production to the measured flux) from that blazar not overproduce the extrapolation, setting an upper limit on the EBL intensity. Other constraints involve placing theoretical limits on the maximum allowed hardness of the deabsorbed spectrum. Recent analysis [28] leaves ~20% uncertainty on the 0.1 – 1000μ EBL intensity.

### 3. The magnetic field in intergalactic space

A field that has been absent in our discussion of intergalactic fields is the magnetic field in intergalactic space, characterized most simply by a magnetic field of average strength $B_{IGMF}$. Energetically, it is small, with an energy density $u_{BIGMF} < 10^{-19} B^2(nG)$ erg/cm$^3$—a magnetic field of order nG is the maximum value allowed by Faraday rotation measurements [29] —but knowledge of it and the correlation length $\lambda_{co}$ over which the magnetic field direction at a random location changes, on average, by 90°—are essential to calculate trajectories of UHECRs. Until recently, knowledge of the value of $B_{IGMF}$ in the voids of intergalactic space was uncertain to 20 orders of magnitude. The importance of the voids in space is that they are where pristine material left over from the big bang and the brief era of nucleosynthesis that followed is found, and indeed this is also where the remnant magnetic field that might be left over from an earlier decoupling phase should be found. The use of γ-ray, radio, and UHECR astronomy to explore mechanisms for magnetogenesis is reviewed in [30].

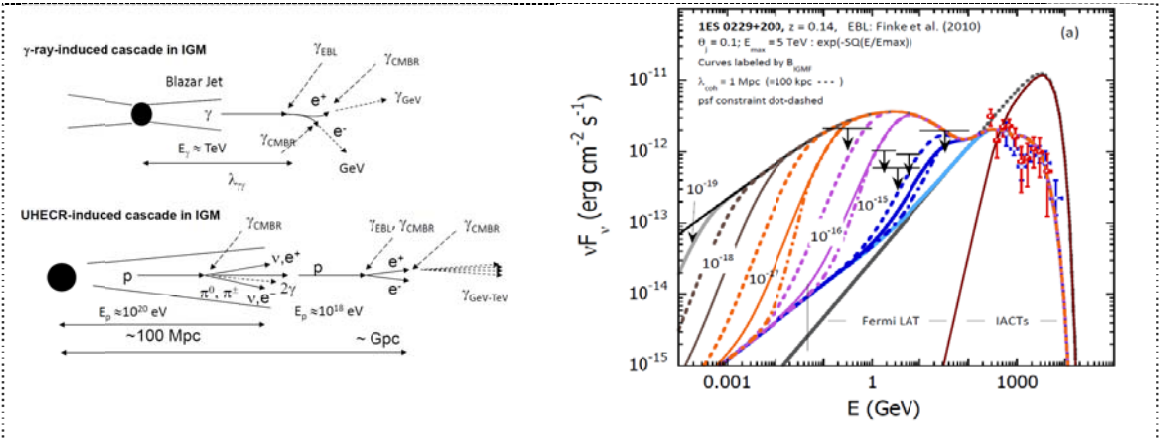

**Figure 5.** Measuring the magnetic field $B_{IGMF}$ of intergalactic space [31]. *Left, upper:* Cartoon showing a γ ray from a TeV blazar source being attenuated by EBL photons, producing an $e^+$-$e^-$ pair that is deflected by $B_{IGMF}$ while scattering CMBR photons to GeV energies. *Right:* Calculations [32] of the reprocessed VHE radiation of the TeV blazar 1ES 0229+200 (z = 0.14) into the GeV range, for different values of $B_{IGMF}$ and $\lambda_{co}$. *Left, lower:* Cartoon of additional high-energy radiation made by UHECR-induced cascades, assuming blazars are sources of UHECRs that travel straight in the IGM.

Fig. 5, upper left, illustrates a process that has received much attention in γ-ray astronomy over the last 20 years: the physics of high-energy TeV γ rays pair-producing with EBL photons (on ~100 Mpc – Gpc length scales determined by the EBL intensity and photon energy $E_\gamma$) to make secondary pairs with Lorentz factors $\gamma = 10^6 \gamma_6 \approx E_\gamma/2m_e \approx 10^6 E_\gamma(\mathrm{TeV})$. The distance such a lepton would travel before losing significant energy by Compton-scattering photons of the CMBR photons is $\lambda_C = 3m_e/4\sigma_T u_{CMBR}\gamma \cong 0.75/\gamma_6$ Mpc, so not so far on intergalactic scales. By comparison, the deflection distance of an $e^+$ or

$e^-$ out of the beam of a relativistic jet has to be smaller than the Larmor radius $r_L = E/QB = m_e\gamma/eB_{IGMF}$ $\cong 0.5\ \gamma_6/B_{-15}$ Mpc, where $B_{-15} = B_{IGMF}/10^{-15}$ G. The features of the reprocessed radiation potentially appear in at least 3 ways: through temporal echoes, angular pair halos, and photon energy spectral features. Much recent interest using both Fermi-LAT and the ground-based γ-ray telescopes has been made to search for angular and temporal features of pair reprocessing of TeV γ rays, though the spectral technique has so far been most successful, as we now outline.

The energy $\varepsilon_s$, in $m_e$ units, of the photons scattered to γ-ray energies is $\varepsilon_s \approx 10^{12}\gamma_6^2 \times 10^{-9} \approx 10^3\ \gamma_6^2$, or $\approx 0.5\ [E_\gamma(TeV)]^2$ GeV, right in the Fermi-LAT range. The absence of such reprocessed emission in the GeV range, as observed in 1ES 0229+200 and a few other blazars with $0.1 <\sim z <\sim 0.2$, requires a sufficiently strong intergalactic magnetic field, $>\sim 10^{-15}$ G by the estimate above, to deflect the pairs away from the observer. Model calculations [32] shown in Fig. 5, right, illustrate a semi-analytic calculation of a single generation cascade, starting with the VHE VERITAS (red) and HESS (blue) data for the TeV blazar 1ES 0229+200, deabsorbing it with EBL model [27], and reprocessing the radiation according to the following equation:

$$\nu F_\nu(\varepsilon_s;t) = \frac{3}{2}(\frac{\varepsilon_s}{\varepsilon_0})^2 \int_{\gamma_{min}}^\infty d\gamma\ \gamma^{-4}(1 - \frac{\varepsilon_s}{4\gamma^2\varepsilon_0}) \int_\gamma^\infty d\gamma_1 \frac{\nu F_\nu\{\exp[\tau_{\gamma\gamma}(\varepsilon;z)]-1\}}{\varepsilon^2},\quad \varepsilon = 2\gamma_1. \quad (3)$$

This equation can be read from right to left as taking the deabsorbed $\nu F_\nu$ energy flux, turning it into one $e^+$ and one $e^-$, each with half the energy of the primary photon, and scattering the CMBR as described by a mono-chromatic radiation field with photon energy $\varepsilon_0 \approx 10^{-9}$ using a simple Thomson scattering kernel [12]. The $\gamma_{min}$ quantity sets a minimum Lorentz factor based on a kinematic limit, a value of γ below which the leptons are deflected out of the line-of-sight (which depends on γ, $B_{-15}$, and $\lambda_{coh}$), and a value limited by the activity time of the TeV blazar.

By examining the right panel of Fig. 5, which is a model [32] of 1ES 0229+200 assuming that it has making TeV γ rays for an arbitrarily long time, one sees that $B_{IGMF}$ must be greater than $\approx 10^{-15}$ G for $\lambda_{coh} \sim 0.1 - 1$ Mpc [31,33]. If the blazar had only been operating for the 3 or so years rather than indefinitely, the lower limit on $B_{IGMF}$ is $2 - 3$ orders of magnitude weaker [32,34]. Uncertainties in the >>TeV part of the blazar spectra used for these studies, which introduces a major uncertainty into the modeling, will be greatly reduced and model limits greatly improved when CTA starts to take data.

## 4. Blazars and radio galaxies in the multi-messenger era

Interest in blazars from the multi-messenger perspective is that they are potential UHECR and high-energy neutrino sources. Blazars are probably not gravitational wave sources for a-LIGO or a-Virgo given the very different size scales and frequencies probed (see the TAUP presentation by Dr. M. Branchesi). A pathway to rapidly spinning supermassive black holes that could power a blazar and produce detectable gravitational radiation is provided by supermassive black-hole coalescence, which would produce events detectable with a space-based LISA-type instrument. (LISA Pathfinder was launched 3 December 2015.) Such events would however be quite rare at LISA sensitivity.

Nevertheless, the existence of radio galaxies and blazars—supermassive black holes with relativistic jets that make highly beamed γ radiation reaching large apparent isotropic γ-ray luminosities $L_{\gamma,iso}$—could help solve the problem of the origin of UHECRs and the TeV – PeV neutrinos. Acceleration of protons through Fermi processes to $>10^{20}$ eV requires isotropic jet powers $P_j$ ($> L_{\gamma,iso}) >\sim 10^{45}$ erg/s (e.g., [35]), from Hillas-condition-type ($r_L <$ size scale of system) considerations. Most blazars achieve this limit, as the high-luminosity, strong-lined FSRQs have $10^{46} <\sim L_{\gamma,iso}$ (erg/s) $<\sim 10^{50}$, whereas the lower-luminosity, weak-lined BL Lac class of blazars have $10^{44} <\sim L_{\gamma,iso}$ (erg/s) $<\sim 10^{47}$ [36].

Not only is there a requirement on apparent jet power to be able to accelerate particles to ulta-high energies, but any source class must also provide sufficient emissivity, $\approx 10^{44}$ erg/Mpc$^3$-yr from Eq. (1), to power the UHECRs or high-energy neutrinos. Although FSRQs and their misaligned counterparts, the luminous Fanaroff-Riley 2 radio galaxies, are individually powerful, their local space density is so low that none are found in the ≈ 100 – 200 Mpc GZK radius, within which > $10^{20}$ eV UHECR sources must be found. More probable is that BL Lac objects and FR-1 radio galaxies like Cen A at 3.5 Mpc are the sources of the UHECRs, given their larger local emissivity, >~ $10^{45}$ erg/Mpc³-yr in 0.1 – 100 GeV γ rays [37]. See [38] for a review of blazars from the γ-ray perspective, and [39] for a review of AGN as neutrino sources, a subject we now consider.

## 5. Origin of high-energy neutrinos and UHECRs

We conclude this paper with a brief discussion of the twin puzzles of the origin of the UHECRs, and the origin of the IceCube excess TeV – PeV neutrinos. These two puzzles may be linked or separate, but any 100 EeV UHECR source is necessarily a candidate source for <~ few PeV ν. The IceCube ν results were well reviewed at this conference by Dr. A. Karle, and besides the broad sky distribution inconsistent with a purely Galactic origin, the other (very) important point to note here is that the excess ≈ 30 TeV – PeV neutrinos are at the Waxman-Bahcall intensity (see Figs. 1 and 2).

*5.1. Neutrinos from blazars and star-forming galaxies*

From Fig. 3, the spectral intensity of the Fermi-LAT EGB can be integrated to give the integral >100 MeV Fermi-LAT EGB intensity $I_γ$(>100 MeV) ≈ 4.4×$10^{-6}$ GeV/cm²-s-sr [15]. By comparison, the Waxman-Bahcall intensity is ≈ 1.5×$10^{-8}$ GeV/cm²-s-sr, with integral >~30 TeV neutrino energy flux a factor ≈ 50 below the fraction of the EGB energy flux emitted by blazars. Thus, only a few % or so of the blazar energy that emerges from the jet as γ rays needs to appear in the form of neutrinos to account for the IceCube observations. SEDs of candidate BL Lac sources of some of the IceCube neutrinos, including both the energy flux in γ rays and ν, are presented in [40,41] to argue in support of blazar (and pulsar wind nebula) association.

To minimize the power to produce the neutrinos, blazar (or GRB) neutrinos would most likely be formed by photohadronic (p + γ → N + π; N is a nucleon ) rather than photo-nuclear (of the type p + p → N + π) processes [42]. A promising mechanism for the origin of blazar neutrinos is photopion production by ultrarelativistic protons interacting with target BLR photons [43]. The BLR can be approximated as a monochromatic isotropic radiation field (in the black-hole frame) with energy $ε_0 ≅ 2×10^{-5}$ of its dominant Ly α line.

The effect of a relativistic boost on a turbulent plasma that entrains relativistic protons with a random pitch angle distribution is to increase the mean stationary (black-hole) frame proton Lorentz factor $γ_p ≅ δ_D γ_p'$ by the Doppler factor of the jet. The threshold condition for making photopions and therefore pionic neutrinos is simply $γ_p ε_0$ >~ $2(m_π/m_e)$, so that the minimum neutrino energy in the black-hole frame is $E_{ν,min} ≅ 0.05\, m_p γ_{p,min} ≅ 1$ PeV, irrespective of the Doppler factor, so long as the proton spectrum is continuous over a large range of energies. The efficiency of extracting energy from the above-threshold protons traveling through the BLR is typically several % [44,45]. By contrast, the ν flux that is produced when the jet protons interact with target internal synchrotron photons is highly sensitive to $δ_D$. Neutrino production in blazars following the blazar sequence was explored in [44] to account for the >~ PeV neutrinos observed with IceCube. In this scenario, a second component is needed to account for the < PeV neutrinos. BL Lac models for ν production can also employ a spine-sheath structure to generalize the available target radiation fields and change threshold behaviors [46].

From a spectral modeling point of view, there is not so much need for hadronic models, inasmuch as purely leptonic models for the broad synchrotron, synchrotron self-Compton, and external Compton components in a relativistic jet generally do a pretty good job of fitting contemporaneous SED data [38]. The requirement of a second ν component in a blazar model for PeV neutrinos is furthermore inelegant, and the predicted structure through the 500 GeV – 1 PeV regime due to the superposition of two components is predicted but observationally unclear. Other criticisms of the blazar scenario is that the cosmic-ray spectrum has to be soft to account for the lack of neutrinos at multi-PeV energies, which would make it difficult for blazars to simultaneously be sources of the UHECRs. The IceCube ν spectrum is fit with $\propto E_\nu^{-2.4}$ spectrum, but BL Lac models for neutrinos show rising spectra at PeV energies [47,48].

A second, low-energy ν component reaching to a few hundred TeV would naturally result from ν production by cosmic rays in star forming galaxies. If they are to produce the IceCube ν, however, cosmic rays with such a steep, −2.7 spectrum like in the Milky Way would overproduce the γ-ray background [49]. A subclass of star-forming galaxies, for example, extreme starburst or ultra-luminous infrared galaxies, or even the compact nuclei of blazars, might accelerate a hard cosmic-ray component that traps <~10 PeV cosmic rays and calorimetrically extracts their energy [18]. The high-efficiency of neutrino production in hadro-nuclear cosmic-ray interactions makes it possible that a subclass of extreme star-forming galaxies make the IceCube neutrinos.

*5.2. UHECRs from blazars*

One of the most fruitful ideas in astro-particle/multi-messenger theory in the last 10 years or so is that UHECRs escaping from blazars can make observable γ-ray and ν signatures [50,51]. Referring to the lower panel of Fig. 5, escaping UHECR protons with energies from $\approx 10^{18}$ eV – few $\times 10^{19}$ eV can travel ≈ Gpc distances (a z ≈ 0.15 source is about 1 Gpc distant), during which time they lose a significant fraction of their energy to Bethe-Heitler pair production (the photo-pion ν-producing reactions for UHECR protons with $E_p >\sim 4 \times 10^{19}$ eV generally occur within a few hundred Mpc of the source). The secondary pairs are made at ultra-high energies, and initiate a Compton/pair-production cascade that emerges with an energy at about the γ-ray horizon energy $E_{\gamma,horizon}$ ($\approx$ 100 GeV/z ; 0.1 <~ z <~ 3) of a source at redshift z.

The cascading Bethe-Heitler pairs make steady—on human timescales—VHE emission, contrary to the notion that blazars are defined as sources that display rapid variability at many frequencies, such as the well-known hyper-variable TeV blazars Mrk 421, Mrk 501, and PKS 2155-304 [38]. The existence of a steady VHE emission signature would be a marker of an UHECR proton-induced cascade, and would void many of the techniques used to constrain the EBL and $B_{IGMF}$. Specifically, if a hard TeV component can be made by UHECRs, the deabsorbed blazar spectra is hardly constrained by its spectral index or by the extension of the Fermi-LAT flux into the VHE range. An extra UHECR-induced VHE component would also mean that the GeV to TeV flux ratio is determined by UHECR processes, not EBL deabsorption and reprocessing, as is assumed in the method to place γ-ray constraints on $B_{IGMF}$.

Remarkably, there is such a class of TeV BL Lac blazars that displays very mild if any variability, and are furthermore located at 0.1 <~ z <~ 0.2. One of these is 1ES 0229+200, which we have already encountered with respect to its use to measure $B_{IGMF}$, and another is 1ES 1101-232 at z = 0.186, and a third is 1ES 0347-121 at z = 0.188. These are favored sources for EBL and $B_{IGMF}$ studies [31,33] yet, because of the possibility that an UHECR component confuses our attempts to constrain the EBL and BIGMF, become less useful than highly variable blazars to constrain EBL and $B_{IGMF}$.

It is therefore urgent to find a way to determine if blazars have additional UHECR-induced components. We [52-54] have been pursuing two avenues to answer whether this is the case. The first is by spectral modeling. By comparing data with models of γ-ray induced and UHECR-induced cascades, which will become especially feasible with CTA, we should be able to answer whether emission from nearby blazars like 1ES 0229+200 [52], and more distant blazars like KUV 00311-1938 at z ≈ 0.61 [53], show evidence for the presence of UHECRs (see Fig. 6).

Although UHECRs made in a blazar jet will escape with a beaming corresponding to the Doppler factor of the inner jet from which they escape, the UHECRs can be rapidly isotropized by deflections in the magnetic fields of the jet core, in the inner galaxy, or in the galaxy cluster field and the structured regions in the vicinity of the galaxy that hosts the blazar. In the worst case for detecting neutrinos, the UHECRs are completely isotropized before escaping into the IGM, and in the best case, they pass rectilinearly through the IGM towards the observer.

If the blazar emission is formed by UHECR-induced cascades, then fitting to the VHE flux for a given proton spectrum with different indices and cutoff energies will likewise make a prediction for the neutrino flux from that blazar. The question of whether UHECRs can travel through the IGM rectilinearly, as needed for the Essey-Kusenko model [50,51] to work, and still obey anisotropy constraints at $\approx 10^{19}$ eV, is addressed in [54]. The hypothesized ~2 Mpc magnetized halo around the Milky Way, required to isotropize the arrival directions of UHECR protons from sources like 1ES 0229+200, if a common feature of all $L_*$-spiral galaxies like the Milky Way, can make it difficult for UHECR protons to travel over Gpc distances without being deflected out of the beam. UHECRs in FSRQs, though unlikely to be the principal source of UHECRs, could still make observable effects through processes in the inner jet (e.g., [55]).

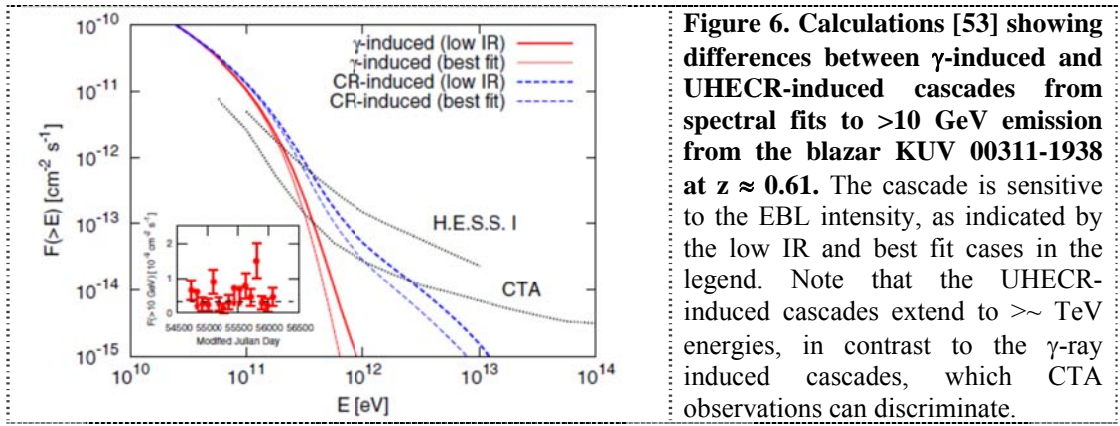

**Figure 6. Calculations [53] showing differences between γ-induced and UHECR-induced cascades from spectral fits to >10 GeV emission from the blazar KUV 00311-1938 at z ≈ 0.61.** The cascade is sensitive to the EBL intensity, as indicated by the low IR and best fit cases in the legend. Note that the UHECR-induced cascades extend to >~ TeV energies, in contrast to the γ-ray induced cascades, which CTA observations can discriminate.

*5.3. Summary*

These are exciting times for multi-messenger astronomy. Our picture of the GeV and TeV γ-ray sky becomes sharper each year. The extragalactic high-energy neutrino window is now open, though ν-astronomy remains no closer to identifying neutrino sources than Auger or TA has for UHECRs. Gravitational-wave astronomy has just detected its first cosmic event.

At the September 2015 Torino TAUP meeting I described the extragalactic night sky at the birth of the multi-messenger era, focusing mainly on blazars as γ-ray, ν, and UHECR sources. Undoubtedly the truth about UHECR and ν origin is more subtle and probably entirely different. But progress in this field is so rapid that it's possible we will know the answers before we know it.

Acknowledgments. I would like to thank Jonathan Biteau and David Williams for comments and the use of Fig. 4, right, from [27], and Marco Ajello for the use of Fig. 3, from [18]. I would like to extend my sincere thanks to Prof. Nicolao Fornengo and Dr. Luca Latronica for the opportunity to attend the XIV International Conference on Topics in Astroparticle and Underground Physics conference in beautiful Torino, Italy.